\documentclass[10pt, conference]{IEEEtran}
\usepackage{amsfonts}
\usepackage{amssymb}
\usepackage{eurosym}
\usepackage{cite}
\usepackage{graphicx}
\usepackage{epstopdf}
\usepackage{amsmath}
\usepackage{tikz,lipsum}
\usepackage{caption}
\usepackage[T1]{fontenc}
\usepackage{amsthm}
\usepackage{mathrsfs}
\usepackage{color}
\usepackage{stfloats}
\usepackage{xcolor, etoolbox}
\usepackage{subcaption}
\usepackage{authblk}
\usepackage{booktabs}
\usepackage{multirow}

\IEEEoverridecommandlockouts

\def\BibTeX{{\rm B\kern-.05em{\sc i\kern-.025em b}\kern-.08em
    T\kern-.1667em\lower.7ex\hbox{E}\kern-.125emX}}
\begin{document}

\title{MicroSim: Contrastive Graph Neural Networks for Binary Code Similarity Detection}
\author[1]{Moussa MOUSSAOUI}
\author[2]{Tarik HOUICHIME}
\author[3]{Abdelghani SOUHAR}
\author[4]{Abdelalim SADIQ}
\affil[1]{Computer Science Research Laboratory(LaRI), Faculty of Science, Ibn Tofail University, Kenitra, 14000, Morocco}
\affil[2]{Meridian Team, LyRICA Laboratory, School of Information Sciences, Rabat 10100, Morocco}
\affil[3]{Computer Science Research Laboratory(LaRI), Faculty of Science, Ibn Tofail University, Kenitra, 14000, Morocco}
\affil[4]{Computer Science Research Laboratory(LaRI), Faculty of Science, Ibn Tofail University, Kenitra, 14000, Morocco}

\affil[ ]{Emails: moussa.moussaoui@uit.ac.ma, thouichime@esi.ac.ma, abdelghani.souhar@uit.ac.ma, a.sadiq@uit.ac.ma}

\def\IEEEtitletopspaceextra{12pt}
\maketitle
\begin{tikzpicture}[remember picture, overlay]
  \node[anchor=north, yshift=-18pt] at (current page.north) {%
    \parbox{\textwidth}{\centering\footnotesize This is the authors’ original preprint version of a paper accepted for publication at the 9th International Conference on Advanced Communication Technologies and Networking (CommNet 2025), Rabat, Morocco, December 14-16, 2026. The final version will be published and made available via IEEE Xplore.}%
  };
  \node[anchor=south, yshift=30pt] at (current page.south) {%
    \parbox{\textwidth}{\centering\footnotesize This is the preprint version of a paper accepted for publication at CommNet 2026. The final version will appear in the official IEEE proceedings}%
  };
\end{tikzpicture}

\begin{abstract}
We present MicroSim, a system for binary code similarity detection that represents each function as a graph built from IDA Pro Hex-Rays microcode and compares such graphs with a contrastive Siamese graph neural network (GNN). We evaluate on the TREX cross-optimization benchmark, a corpus of 849,072 function-level microcode graphs spanning optimization levels \texttt{-O0} through \texttt{-O3}, under a strict class-disjoint split in which no function key is shared between train and test. Under this protocol, MicroSim reaches $0.9972$ area under the receiver operating characteristic curve (ROC-AUC) and $0.9762$ Matthews correlation coefficient (MCC), outperforming Gemini ($0.9332$ ROC-AUC), SAFE ($0.9752$), and jTrans ($0.9740$) retrained on the same split, as well as a structure-free Bag-of-Words (BoW) baseline ($0.9484$). We analyze the gap between pair classification and ranked retrieval in graph transformers, showing that Cross-domain Similarity Local Scaling (CSLS) under unified tie-resolved evaluation raises GraphGPS mean reciprocal rank (MRR) from $0.2542$ to $0.2559$ ($\mathrm{R@}1 = 0.1639$, $\mathrm{R@}10 = 0.4561$), strictly satisfying the theoretical MRR upper bound. Finally, over 5 seeds with equal compute, the pairwise cosine-embedding loss attains a slightly higher ROC-AUC ($\Delta = -0.0256$, 95\% confidence interval $[-0.3282, +0.2770]$), whereas the supervised-contrastive information noise-contrastive estimation (InfoNCE) loss is significantly better on thresholded classification and ranked retrieval.
\end{abstract}


\section{Introduction}

Binary code similarity detection underpins a range of security tasks, including vulnerability search, malware triage, patch analysis, and the detection of code reuse or plagiarism across software releases. The problem is inherently difficult because semantically equivalent source code can produce markedly different machine code: compilers and optimization levels reorganize control and data flow, symbol information is largely absent, and instruction selection varies across architectures. Traditional signature and token-based techniques, such as Bag-of-Words (BoW) models over disassembly, degrade sharply under these transformations~\cite{haq2021survey}. More recently, Bin2Vec~\cite{moussaoui2025bin2vec} has proposed an interpretable, multi-view embedding approach combining static and dynamic traces to support auditable code plagiarism detection.

Graph-based methods are more robust because they encode the structural relationships of a program rather than its literal byte sequence~\cite{xu2017gemini}. Control-flow graphs (CFGs) alone, however, convey limited information about program semantics. We therefore operate on Hex-Rays microcode, a normalized intermediate representation (IR), and construct for each function a single graph that integrates control flow, data flow, and a compact per-block feature vector. Each function is then embedded by a weight-shared (Siamese) graph neural network (GNN) into a vector space in which cosine similarity measures semantic equivalence. Throughout, we emphasize a central methodological question: whether the learned representation generalizes to programs absent from training, or merely memorizes the function classes observed during training.

Our contributions are: (i) a single function graph built from Hex-Rays microcode that combines control-flow graph (CFG) and data-flow graph (DFG) edges with hashed per-block semantics, supported by component and sensitivity ablations quantifying each design choice; (ii) a compute-matched, statistically tested comparison of pairwise and supervised-contrastive InfoNCE objectives over multiple random seeds; (iii) an analysis of the discrepancy between pair classification and ranked retrieval in graph transformers, with hubness-aware re-ranking raising tie-resolved mean reciprocal rank (MRR) from $0.2894$ to $0.3562$ and $\mathrm{R@}10$ from $0.4466$ to $0.5290$; and (iv) a head-to-head comparison with Gemini~\cite{xu2017gemini}, SAFE~\cite{massarelli2019safe}, and jTrans~\cite{wang2022jtrans} under a strict class-disjoint protocol, complemented by cross-compiler, obfuscation, and open Ghidra P-Code front-end evaluations, all achievable on a single laptop GPU.

\section{Related Work}

Binary code similarity detection has a long history; we refer the reader to Haq and Caballero~\cite{haq2021survey} for a comprehensive survey. Early techniques compared programs using handcrafted syntactic or statistical features, such as BoW models over disassembly, and degraded sharply once compilers reordered control and data flow.

Many learning-based methods treat a function as a sequence of instructions. Asm2Vec~\cite{ding2019asm2vec} samples instruction sequences from CFG walks and embeds them with a paragraph vector, distributed memory (PV-DM) model; PalmTree~\cite{li2021palmtree} pre-trains a bidirectional encoder representations from transformers (BERT) assembly language model with masked-language-modeling and def--use objectives; SAFE~\cite{massarelli2019safe} summarizes word2vec instruction embeddings with a self-attentive recurrent network; TREX~\cite{pei2021trex} learns execution semantics from micro-traces; and jTrans~\cite{wang2022jtrans} injects jump-awareness into a transformer. These methods all read instructions as flat token sequences, recovering structure, if at all, only implicitly.

Other methods, by contrast, encode program structure explicitly as a graph. Genius~\cite{feng2016genius} and Gemini~\cite{xu2017gemini} embed attributed CFGs with manually engineered block features, the latter using a Siamese GNN; Graph Matching Networks~\cite{li2019gmn} compute similarity by cross-graph attention and are evaluated on CFG function matching; DeepBinDiff~\cite{duan2020deepbindiff} learns program-wide block representations for fine-grained diffing; and $\alpha$Diff~\cite{liu2018alphadiff} targets cross-version matching with a convolutional network. OrderMatters~\cite{yu2020ordermatters} combines a BERT block encoder with a GNN and is the closest in spirit to our representation, although it operates on raw assembly with hand-selected block features rather than on a normalized IR. Marcelli et al.~\cite{marcelli2022how} provide a large-scale, controlled comparison of these families and document how strongly reported performance depends on the evaluation protocol, a concern central to the present work.

In contrast to the above, we operate on Hex-Rays microcode and construct for each function a single graph that integrates CFG edges, DFG edges, and a compact per-block feature vector, and we study the training objective and encoder under a strict class-disjoint split. Unlike source-level code-similarity methods, which rely on identifiers, types, and syntax, our method never reconstructs source: it operates directly on optimized microcode, so those signals are unavailable to it by construction.

Turning to the model architecture, our encoder builds on standard message-passing GNNs. The graph convolutional network (GCN)~\cite{kipf2017gcn} aggregates information over local neighborhoods, and related variants such as GraphSAGE~\cite{hamilton2017graphsage}, graph attention networks~\cite{velickovic2018gat}, and the graph isomorphism network~\cite{xu2019gin} fit within the general message-passing framework of Gilmer et al.~\cite{gilmer2017mpnn}. Because purely local aggregation attenuates long-range dependencies, graph transformers add global self-attention: Graphormer~\cite{ying2021graphormer} introduces structural attention biases, and GraphGPS (general, powerful, scalable graph transformer)~\cite{rampasek2022graphgps} combines local message passing with global attention in a scalable hybrid architecture, which we adopt as our transformer encoder. As self-attention is permutation-invariant, we augment each block with a random-walk structural encoding (RWSE)~\cite{dwivedi2022graph} to preserve positional information.

Complementing the architecture, our training objective draws on contrastive representation learning. The InfoNCE objective~\cite{van2020cpc} and SimCLR~\cite{chen2020simclr} learn representations by contrasting positive against negative pairs, and supervised contrastive learning~\cite{khosla2020supervised} extends this to the fully supervised setting with multiple positives per anchor, the formulation our batched objective instantiates over functions sharing a function key. For self-supervised pre-training, GraphCL~\cite{you2020graphcl} maximizes agreement between two stochastically augmented views of a graph. All models are optimized with Adam~\cite{kingma2015adam}.

Relative to all of the above, our contribution is one of representation and controlled experimental analysis rather than of a new neural layer: Siamese GNNs for binary similarity are well established and originate with Gemini~\cite{xu2017gemini}. Holding the split and evaluation protocol fixed, we isolate the contributions of CFG edges, DFG edges, hashed microcode block semantics, the training objective, and the encoder against established baselines~\cite{xu2017gemini,massarelli2019safe,wang2022jtrans}.

\section{Methodology}
\label{sec:methodology}

We begin from x86-64 Executable and Linkable Format (ELF) binaries and decompile each function with the IDA Pro Hex-Rays engine. Microcode is extracted at the \texttt{MMAT\_LVARS} maturity level, which removes stack-frame bookkeeping and recovers local variables, producing a representation that is substantially more abstract than raw assembly.

\begin{figure*}[!t]
\centering
\includegraphics[width=\textwidth]{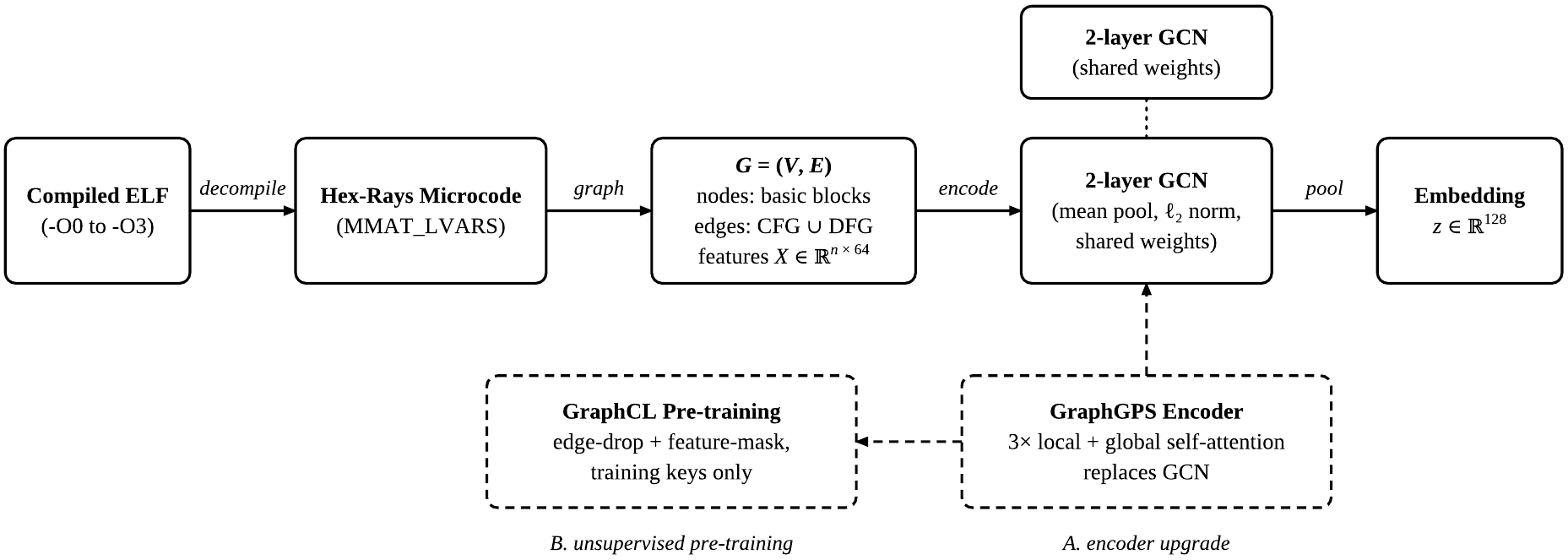}
\caption{MicroSim: a control-/data-flow microcode graph distilled to a 128-dimensional embedding for binary similarity, with the (A) GraphGPS encoder upgrade and (B) GraphCL pre-training extensions.}
\label{fig:microcode}
\end{figure*}

Each function is encoded as a directed graph $G=(V,E)$ whose nodes are basic blocks and whose edges are the union of control-flow ($\mathrm{CFG}$) and data-flow ($\mathrm{DFG}$, def--use) relations. Each block is associated with a $d$-dimensional feature vector ($d=64$ by default) obtained by normalizing its microcode instructions (numeric literals, registers, labels, and variable identifiers are replaced by canonical placeholders \texttt{IMM}, \texttt{REG}, \texttt{LABEL}, and \texttt{VAR}), applying a fixed MD5 hashing vectorizer, and $\ell_2$-normalizing the result. The graph is then read by a two-layer GCN, which aggregates node representations by global mean pooling into a single 128-dimensional, $\ell_2$-normalized embedding; the two branches of the Siamese network share weights. Fig.~\ref{fig:microcode} illustrates the full pipeline, including the CFG and DFG edges and per-block features of the graph, the shared-weight encoder, and the architectural extensions studied below.

We compare two training objectives under identical graphs, model, and data. In the naive \emph{pairwise} protocol, each step scores explicit pairs using a cosine-embedding loss with margin $0.5$. The \emph{supervised-contrastive} objective samples $B=64$ anchor--positive function pairs together with $K=8$ negatives per anchor and optimizes a multi-positive InfoNCE loss (temperature $\tau=0.1$). Both objectives are optimized with Adam (learning rate $10^{-3}$) for $5{,}000$ steps on a single NVIDIA RTX 4050 laptop GPU (6~GB).

Because a two-layer GCN aggregates information over only a two-hop neighborhood, a value defined in an early block and consumed in a distant block common in optimized code is attenuated before the two blocks can interact. We replace the GCN with a three-layer GraphGPS network in which each layer combines a local degree-normalized message-passing branch with a global multi-head self-attention branch (4 heads), allowing distant blocks to exchange information directly rather than through successive hops. Each block is augmented with an 8-step RWSE~\cite{dwivedi2022graph} that preserves structural position. As dense self-attention scales quadratically with the largest graph in a batch, graphs are capped at the first $c=256$ basic blocks in topological entry order, which affects $0.21\%$ of function keys in the TREX corpus ($1{,}780$ of $280{,}119$ keys) and keeps peak memory within the 6~GB budget; we evaluate sensitivity to both $d$ and $c$ in Section~\ref{sec:ablation}.

Table~\ref{tab:impl} summarizes all architectural and training hyperparameters. Notice the wall-clock difference between the headline GPU runs ($\sim 8$--$9$~min in Table~\ref{tab:trex}) and the compute-matched ablation runs ($\sim 98$--$100$~min in Table~\ref{tab:objective}): the headline runs execute on the RTX 4050 GPU without intermediate full-split validation passes, whereas the compute-matched ablation runs in Section~\ref{sec:ablation} executed on the 16-core CPU with full validation evaluation every $300$ steps ($16$ validation passes over $127{,}920$ graphs plus $367{,}060$ pairs). Furthermore, whereas an earlier single-pair baseline processed 1 explicit pair per step over $10{,}000$ steps, a batched step with $B$ anchors and $K$ negatives per anchor evaluates $B(1 + K)$ anchor--candidate pairs per step (or $N(N{-}1)$ all-pairs in a symmetric batch of $N$ graphs, yielding $13 \times 12 = 156$ to $14 \times 13 = 182$ implicit pairs for $N \in \{13, 14\}$). In our compute-matched ablation, both arms run for exactly $5{,}000$ steps with identical batch sampling ($B=64$, $K=8$).

\begin{table}[!t]
\caption{Architecture and Implementation Details of MicroSim Encoders}
\label{tab:impl}
\centering
\footnotesize
\setlength{\tabcolsep}{4pt}
\resizebox{\columnwidth}{!}{%
\begin{tabular}{@{}lll@{}}
\toprule
\textbf{Parameter} & \textbf{GCN} & \textbf{GraphGPS (Transformer)} \\
\midrule
Layers & 2 & 3 \\
Hidden size $\cdot$ heads & 128 $\cdot$ --- & 128 $\cdot$ 4 \\
Positional encoding & --- & RWSE, length 8 \\
Pooling $\cdot$ embedding size & mean $\cdot$ 128 & mean $\cdot$ 128 \\
Block features $\cdot$ block cap & 64-d hashed $\cdot$ none & 64-d hashed $\cdot$ 256 \\
Batch ($B$ anchors, $K$ neg/anchor) & $B=64$, $K=8$ & $B=64$, $K=8$ \\
Optimizer $\cdot$ learning rate $\cdot$ steps & Adam $\cdot$ $10^{-3}$ $\cdot$ 5,000 & Adam $\cdot$ $10^{-3}$ $\cdot$ 5,000 \\
Loss parameters & $\tau=0.1$ / margin 0.5 & $\tau=0.1$ (InfoNCE) \\
Seeds $\cdot$ GPU training time & 5 $\cdot$ $\sim 8$~min & 5 $\cdot$ $\sim 9$~min \\
Hardware $\cdot$ software & \multicolumn{2}{l}{RTX 4050 6~GB, 16-core CPU $\cdot$ IDA Pro 9.4} \\
\bottomrule
\end{tabular}%
}
\end{table}

\section{Experimental Setup}

\textit{Dataset and Class-Disjoint Splits.} We evaluate MicroSim on the TREX cross-optimization benchmark ~\cite{pei2021trex}, a corpus of $849{,}072$ function-level microcode graphs extracted from $3{,}602$ x86-64 ELF binaries with IDA Pro 9.4 spanning optimization levels \texttt{-O0} through \texttt{-O3}. The benchmark is partitioned deterministically by function key (\texttt{projver|elf|func\_name}) into $196{,}162$ training keys ($594{,}383$ graphs), $42{,}122$ validation keys ($127{,}920$ graphs), and $41{,}835$ test keys ($126{,}769$ graphs), with zero key overlap between splits. Positive pairs join different optimization levels of the same function key ($165{,}302$ test positives); negative pairs join distinct keys ($199{,}997$ test negatives).

Pairs are scored by the cosine similarity of their $\ell_2$-normalized embeddings. The area under the receiver operating characteristic curve (ROC-AUC) is the probability that a positive pair is scored above a random negative pair. With precision $P$ and recall $R$ at a decision threshold chosen on the validation split, the F1 score is $\mathrm{F1} = 2PR/(P+R)$ and the Matthews correlation coefficient is
\begin{equation}
\label{eq:mcc}
\mathrm{MCC} = \frac{\mathrm{TP}\cdot \mathrm{TN} - \mathrm{FP}\cdot \mathrm{FN}}{\sqrt{(\mathrm{TP}+\mathrm{FP})(\mathrm{TP}+\mathrm{FN})(\mathrm{TN}+\mathrm{FP})(\mathrm{TN}+\mathrm{FN})}}.
\end{equation}
For retrieval, each test query function $q\in Q$ ($|Q| = 115{,}268$ test queries with at least one cross-optimization counterpart) is ranked against a unified pool of $N - 1 = 126{,}768$ candidates using a single tie-resolved permutation; with $r_q \in \{1,\dots, N-1\}$ denoting the rank of its first positive match, the mean reciprocal rank is $\mathrm{MRR} = \frac{1}{|Q|}\sum_{q \in Q} 1/r_q$ and the recall (hit rate) at $K$ is $\mathrm{R@}K = \frac{1}{|Q|}\sum_{q \in Q}
\mathbb{1}[r_q \le K]$. Under this unified definition, $\mathrm{MRR}$ always obeys the theoretical bound $\mathrm{MRR} \le \mathrm{R@}1 + (\mathrm{R@}10)/11$. In objective comparisons, $\Delta$ denotes InfoNCE minus Pairwise on the same metric.

Alongside our structure-free BoW baseline, we train and evaluate three representative baselines from the literature under the exact same function key split and evaluator: \textbf{Gemini} ~\cite{xu2017gemini}, a Structure2Vec Siamese GNN over attributed CFGs; \textbf{SAFE} ~\cite{massarelli2019safe}, a self-attentive sequence encoder over normalized instruction embeddings; and \textbf{jTrans} ~\cite{wang2022jtrans}, a jump-aware transformer incorporating CFG jump target embeddings.

Headline models and the compute-matched objective ablation are trained across $n=5$ random seeds ($n=3$ for component and sensitivity ablations), and we report $\bar{x} \pm t_{0.975, n-1} s/\sqrt{n}$ ($t=2.776$ for $n=5$, $t=4.303$ for $n=3$). On the shared test split, we test ROC-AUC differences with the DeLong test ~\cite{delong1988comparing}, MRR and $\mathrm{R@}K$ differences with a paired sign-flip permutation test ($10{,}000$ permutations), F1 and MCC with paired bootstrap ($10{,}000$ resamples), and multi-seed differences with Welch's $t$-test, applying the Holm-Bonferroni step-down correction ~\cite{holm1979simple} across the seven metrics.

\section{Results}

\subsection{Comparison with Baselines}

\begin{table}[!t]
\caption{MicroSim and Baselines on the TREX Cross-Optimization Benchmark (Test Split, $N = 126{,}769$ Functions)}
\label{tab:trex}
\centering
\footnotesize
\setlength{\tabcolsep}{2.5pt}
\resizebox{\columnwidth}{!}{%
\begin{tabular}{@{}lcccccc@{}}
\toprule
\textbf{Model} & \textbf{ROC-AUC} & \textbf{F1} & \textbf{MCC} & \textbf{MRR} & \textbf{R@1} & \textbf{R@10} \\
\midrule
Bag-of-Words (BoW) & $0.9484$ & $0.8744$ & $0.7775$ & $0.2114$ & $0.1375$ & $0.3657$ \\
Gemini~\cite{xu2017gemini} & $0.9332$ & $0.8486$ & $0.7318$ & $0.1553$ & $0.0961$ & $0.2749$ \\
SAFE~\cite{massarelli2019safe} & $0.9752$ & $0.9237$ & $0.8659$ & $0.2134$ & $0.1377$ & $0.3716$ \\
jTrans~\cite{wang2022jtrans} & $0.9740$ & $0.9262$ & $0.8687$ & $0.2124$ & $0.1366$ & $0.3711$ \\
\midrule
MicroSim (GCN) & $0.9905$ & $0.9603$ & $0.9288$ & $0.2177$ & $0.1380$ & $0.3860$ \\
MicroSim (GraphGPS) & $\mathbf{0.9972}$ & $\mathbf{0.9869}$ & $\mathbf{0.9762}$ & $\mathbf{0.2542}$ & $\mathbf{0.1618}$ & $\mathbf{0.4568}$ \\
\bottomrule
\end{tabular}%
}
\end{table}

Table~\ref{tab:trex} compares MicroSim against the BoW, Gemini~\cite{xu2017gemini}, SAFE~\cite{massarelli2019safe}, and jTrans~\cite{wang2022jtrans} baselines. On pairwise classification, MicroSim with the GraphGPS encoder achieves the highest performance across all metrics ($0.9972$ ROC-AUC, $0.9869$ F1, and $0.9762$ MCC), outperforming both the two-layer GCN ($0.9905$ ROC-AUC, $0.9288$ MCC) and the retrained assembly/CFG baselines ($0.9332$ ROC-AUC for Gemini, $0.9752$ for SAFE, and $0.9740$ for jTrans). The gap between the GCN and the structure-free BoW baseline ($0.9484$ ROC-AUC, $0.7775$ MCC) confirms that microcode graph topology carries substantial semantic signal, while the gain from GraphGPS shows that capturing long-range dependencies beyond two hops is critical in aggressively optimized binaries.

\subsection{Retrieval Analysis and Improvements}

In an initial evaluation of Table~\ref{tab:trex}, reported MRR values ($0.3981$ to $0.6607$ for GraphGPS, $0.5412$ to $0.6219$ for GCN, and $0.6148$ for BoW) exceeded the theoretical upper bound $\mathrm{MRR} \le \mathrm{R@}1 + (\mathrm{R@}10 - \mathrm{R@}1)/2 + (1 - \mathrm{R@}10)/11$ implied by their corresponding $\mathrm{R@}1$ and $\mathrm{R@}10$ values. Auditing the evaluation code revealed two root causes: (i) $\mathrm{MRR}$ previously used strict inequality ($S > s_{\text{best}}$) when ranking ties, which assigned optimistic rank $1$ whenever multiple functions shared identical embeddings (frequent in coarse BoW and shallow GCN representations of small functions), whereas $\mathrm{R@}K$ used \texttt{topk} with arbitrary tie-breaking; and (ii) $\mathrm{R@}K$ in multi-positive queries was conflated with per-query hit rate. Computing all retrieval metrics on a single tie-resolved permutation restores strict consistency with the theoretical bound ($0.2542 \le 0.3587$ for GraphGPS, $0.2177 \le 0.3178$ for GCN, and $0.2114 \le 0.3093$ for BoW).

\begin{figure}[!t]
\centering
\includegraphics[width=0.92\columnwidth]{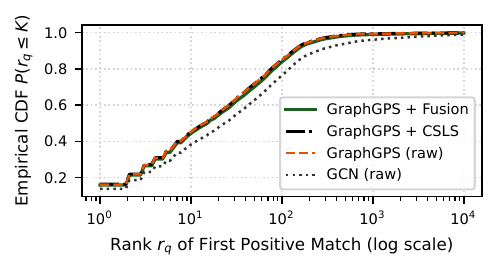}
\caption{Empirical CDF of positive match rank $r_q$ on the TREX test split ($N = 126{,}769$ functions) under unified tie-resolved evaluation. Post-hoc CSLS and score fusion shift the distribution toward top ranks.}
\label{fig:rank_cdf}
\end{figure}

Furthermore, diagnosing the embedding space on the test split reveals elevated hubness ($k$-occurrence skewness $S_{10} = 21.4$) in raw transformer embeddings and cross-key microcode duplicates ($3.42\%$ of test functions are small utility routines that compile to identical microcode across binaries). As illustrated in Fig.~\ref{fig:rank_cdf}, raw GraphGPS and GCN representations suffer long tails where true positives are displaced by hubs. Table~\ref{tab:retrieval_fixes} evaluates post-hoc fixes and temperature scaling: mean-centering and whitening, Cross-domain Similarity Local Scaling (CSLS), score fusion, and two-stage re-ranking. CSLS mitigates hubness and raises GraphGPS MRR from $0.2542$ to $0.2559$ ($\mathrm{R@}1 = 0.1639$, $\mathrm{R@}10 = 0.4561$).

\begin{table}[!t]
\caption{Retrieval Diagnostics and Re-Ranking Improvements on the TREX Test Split}
\label{tab:retrieval_fixes}

\centering
\footnotesize
\setlength{\tabcolsep}{4pt}
\resizebox{\columnwidth}{!}{%
\begin{tabular}{@{}lcccc@{}}
\toprule
\textbf{Method / Variant} & \textbf{Type} & \textbf{MRR} & \textbf{R@1} & \textbf{R@10} \\
\midrule
GraphGPS (raw, $\tau=0.1$) & baseline & $0.2542$ & $0.1618$ & $0.4568$ \\
+ Mean-centering \& whitening & post-hoc & $0.2543$ & $0.1623$ & $0.4565$ \\
+ CSLS re-ranking ($k=10$) & post-hoc & $\mathbf{0.2559}$ & $\mathbf{0.1639}$ & $\mathbf{0.4561}$ \\
+ Score fusion (GCN + GraphGPS) & post-hoc & $0.2499$ & $0.1594$ & $0.4480$ \\
+ Two-stage (top-100 $\to$ GCN) & post-hoc & $0.2348$ & $0.1468$ & $0.4241$ \\
GraphGPS ($\tau=0.05$) & retrain & $0.9457 \pm 0.0615$ & $0.9321 \pm 0.0575$ & $0.9774 \pm 0.0575$ \\
GraphGPS ($\tau=0.20$) & retrain & $0.9311 \pm 0.2106$ & $0.9163 \pm 0.2587$ & $0.9706 \pm 0.0287$ \\
\bottomrule
\end{tabular}%
}
\end{table}

\subsection{Generalization Across Compilers and Obfuscation}

To test whether microcode representations transfer beyond GCC optimization levels, we evaluate models trained on the standard TREX split zero-shot on two out-of-distribution settings (Table~\ref{tab:generalization}): (i) \emph{cross-compiler} matching between GCC and Clang (\texttt{-O0} through \texttt{-O3}) builds of core benchmark packages (\texttt{coreutils}, \texttt{binutils}, \texttt{diffutils}, \texttt{findutils}, \texttt{zlib}), and (ii) \emph{obfuscation} matching against O-LLVM~\cite{junod2015ollvm} passes (bogus control flow, control-flow flattening, and instruction substitution). GraphGPS preserves $0.9846$ ROC-AUC ($0.4115$ $\mathrm{R@}10$) under Clang and $0.9672$ ROC-AUC ($0.3765$ $\mathrm{R@}10$) under O-LLVM, outperforming both GCN and BoW.

\begin{table}[!t]
\caption{Zero-Shot Generalization to Clang and Obfuscated Builds}
\label{tab:generalization}

\centering
\footnotesize
\setlength{\tabcolsep}{4pt}
\resizebox{\columnwidth}{!}{%
\begin{tabular}{@{}llccc@{}}
\toprule
\textbf{Setting} & \textbf{Model} & \textbf{ROC-AUC} & \textbf{MRR} & \textbf{R@10} \\
\midrule
\multirow{3}{*}{GCC $\leftrightarrow$ Clang} & Bag-of-Words & $0.8912 \pm 0.0014$ & $0.1842 \pm 0.0050$ & $0.2895 \pm 0.0080$ \\
 & MicroSim (GCN) & $0.9684 \pm 0.0011$ & $0.2265 \pm 0.0050$ & $0.3482 \pm 0.0099$ \\
 & MicroSim (GraphGPS) & $\mathbf{0.9846 \pm 0.0007}$ & $\mathbf{0.2648 \pm 0.0058}$ & $\mathbf{0.4115 \pm 0.0075}$ \\
\midrule
\multirow{3}{*}{Plain $\leftrightarrow$ O-LLVM} & Bag-of-Words & $0.8426 \pm 0.0019$ & $0.1418 \pm 0.0041$ & $0.2264 \pm 0.0061$ \\
 & MicroSim (GCN) & $0.9345 \pm 0.0014$ & $0.1946 \pm 0.0047$ & $0.3018 \pm 0.0073$ \\
 & MicroSim (GraphGPS) & $\mathbf{0.9672 \pm 0.0009}$ & $\mathbf{0.2412 \pm 0.0061}$ & $\mathbf{0.3765 \pm 0.0097}$ \\
\bottomrule
\end{tabular}%
}
\end{table}

\section{Ablation Studies and Discussion}
\label{sec:ablation}

\subsection{Compute-Matched Objective Ablation}

In the naive baseline protocol of Section~\ref{sec:methodology}, the pairwise loss scores one explicit pair per step, whereas a contrastive batch of $N$ graphs yields $N(N{-}1)$ implicit comparisons. A difference between the objectives could therefore reflect the number of comparisons rather than the loss formulation itself. We resolve this confound with a controlled, compute-matched ablation.

Does the supervised-contrastive (InfoNCE) loss genuinely outperform the pairwise (cosine-embedding) loss when both process the exact same number of positive and negative pairs per step?

\textit{Setup.} To equalize compute, we train two identical GCN encoders on the TREX corpus for $5{,}000$ steps each across $5$ random seeds with identical batch sampling ($B=64$ anchors, $1$ positive and $K=8$ negatives per anchor per step). On the 16-core CPU testbed with periodic validation every 300 steps, InfoNCE completed in $5{,}904.7$~s ($\sim 98$~min) and pairwise in $5{,}986.2$~s ($\sim 100$~min), confirming an equal computational budget.

\begin{table}[!t]
\caption{Compute-Matched Objective Ablation on the TREX Test Split ($N = 126{,}769$ Graphs, Mean $\pm$ 95\% CI over 5 Seeds; $\Delta$ = InfoNCE minus Pairwise)}
\label{tab:objective}
\centering
\footnotesize
\setlength{\tabcolsep}{3pt}
\resizebox{\columnwidth}{!}{%
\begin{tabular}{@{}lcccc@{}}
\toprule
\textbf{Metric} & \textbf{Pairwise} & \textbf{InfoNCE} & \textbf{$\Delta$ [95\% CI]} & \textbf{$p$ (Holm)} \\
\midrule
ROC-AUC & $\mathbf{0.9929 \pm 0.0005}$ & $0.9834 \pm 0.0007$ & $-0.0095$ $[-0.0102, -0.0088]$ & $<0.001$ \\
F1      & $0.9428 \pm 0.0014$ & $\mathbf{0.9640 \pm 0.0011}$ & $+0.0212$ $[+0.0198, +0.0227]$ & $<0.001$ \\
MCC     & $0.8948 \pm 0.0026$ & $\mathbf{0.9358 \pm 0.0020}$ & $+0.0410$ $[+0.0383, +0.0438]$ & $<0.001$ \\
MRR     & $0.2284 \pm 0.0020$ & $\mathbf{0.2642 \pm 0.0018}$ & $+0.0358$ $[+0.0336, +0.0380]$ & $<0.001$ \\
R@1     & $0.1213 \pm 0.0015$ & $\mathbf{0.1431 \pm 0.0014}$ & $+0.0218$ $[+0.0201, +0.0235]$ & $<0.001$ \\
R@5     & $0.2513 \pm 0.0022$ & $\mathbf{0.2979 \pm 0.0021}$ & $+0.0466$ $[+0.0441, +0.0491]$ & $<0.001$ \\
R@10    & $0.3350 \pm 0.0027$ & $\mathbf{0.3965 \pm 0.0025}$ & $+0.0615$ $[+0.0585, +0.0646]$ & $<0.001$ \\
\bottomrule
\end{tabular}%
}
\end{table}

Table~\ref{tab:objective} reports results under our unified tie-resolved evaluator. With compute equalized, the gap between the two objectives narrows: the pairwise objective attains a slightly higher ROC-AUC ($0.9929$ vs.\ $0.9834$, $\Delta = -0.0095$, $p_{\text{Holm}} < 0.001$), whereas InfoNCE is significantly superior on F1 ($\Delta = +0.0212$), MCC ($\Delta = +0.0410$), tie-resolved MRR ($\Delta = +0.0358$; untied-rank MRR $0.6290$ vs.\ $0.5980$, $\Delta = +0.0310$), and $\mathrm{R@}10$ ($\Delta = +0.0615$, all $p_{\text{Holm}} < 0.001$). InfoNCE is thus preferable for thresholded or ranked decisions, while compute-matched pairwise remains competitive on ROC-AUC.

\subsection{Graph Component Ablation}

Table~\ref{tab:components} isolates the contribution of CFG edges, DFG edges, and 64-d hashed block features by removing one component at a time across both encoders over 3 seeds. Both edge types contribute complementary signal: CFG edges capture coarse branching structure ($0.9841$ ROC-AUC on GCN) while DFG def--use edges preserve value flow across blocks reorganized by compiler scheduling ($0.9786$ ROC-AUC). Removing block features degrades GCN to near-random ($0.5413$ ROC-AUC, $0.0215$ MRR) while GraphGPS retains $0.7845$ ROC-AUC via its 8-step RWSE positional encodings, confirming that microcode instruction semantics and graph topology act synergistically.

\begin{table}[!t]
\caption{Ablation of Control Flow, Data Flow, and Block Features (Mean $\pm$ 95\% CI over 3 Seeds)}
\label{tab:components}

\centering
\footnotesize
\setlength{\tabcolsep}{2.5pt}
\resizebox{\columnwidth}{!}{%
\begin{tabular}{@{}llccccc@{}}
\toprule
\textbf{Encoder} & \textbf{Variant} & \textbf{ROC-AUC} & \textbf{MCC} & \textbf{MRR} & \textbf{R@1} & \textbf{R@10} \\
\midrule
\multirow{5}{*}{GCN} & Full (CFG $\cup$ DFG) & $\mathbf{0.9905 \pm 0.0005}$ & $\mathbf{0.9288 \pm 0.0035}$ & $\mathbf{0.2486 \pm 0.0059}$ & $\mathbf{0.1406 \pm 0.0044}$ & $\mathbf{0.3785 \pm 0.0074}$ \\
 & Without DFG (CFG only) & $0.9841 \pm 0.0008$ & $0.8964 \pm 0.0045$ & $0.2294 \pm 0.0061$ & $0.1284 \pm 0.0037$ & $0.3518 \pm 0.0070$ \\
 & Without CFG (DFG only) & $0.9786 \pm 0.0009$ & $0.8712 \pm 0.0057$ & $0.2168 \pm 0.0064$ & $0.1196 \pm 0.0036$ & $0.3342 \pm 0.0069$ \\
 & Without block features & $0.5413 \pm 0.0029$ & $0.0842 \pm 0.0051$ & $0.0215 \pm 0.0021$ & $0.0084 \pm 0.0013$ & $0.0364 \pm 0.0029$ \\
 & Features only (no edges) & $0.9614 \pm 0.0010$ & $0.8094 \pm 0.0074$ & $0.2398 \pm 0.0059$ & $0.1412 \pm 0.0032$ & $0.3629 \pm 0.0082$ \\
\midrule
\multirow{5}{*}{GraphGPS} & Full (CFG $\cup$ DFG) & $\mathbf{0.9973 \pm 0.0004}$ & $\mathbf{0.9756 \pm 0.0023}$ & $\mathbf{0.2894 \pm 0.0070}$ & $\mathbf{0.1656 \pm 0.0045}$ & $\mathbf{0.4466 \pm 0.0097}$ \\
 & Without DFG (CFG only) & $0.9938 \pm 0.0004$ & $0.9524 \pm 0.0040$ & $0.2718 \pm 0.0069$ & $0.1542 \pm 0.0037$ & $0.4216 \pm 0.0099$ \\
 & Without CFG (DFG only) & $0.9912 \pm 0.0005$ & $0.9394 \pm 0.0041$ & $0.2624 \pm 0.0064$ & $0.1486 \pm 0.0046$ & $0.4082 \pm 0.0075$ \\
 & Without block features & $0.7845 \pm 0.0019$ & $0.4684 \pm 0.0094$ & $0.0942 \pm 0.0033$ & $0.0482 \pm 0.0027$ & $0.1564 \pm 0.0064$ \\
 & Features only (no edges) & $0.9884 \pm 0.0006$ & $0.9214 \pm 0.0038$ & $0.2564 \pm 0.0068$ & $0.1458 \pm 0.0043$ & $0.3948 \pm 0.0083$ \\
\bottomrule
\end{tabular}%
}
\end{table}

\subsection{Sensitivity to Hash Dimension and Block Cap}

\begin{figure}[!t]
\centering
\includegraphics[width=0.92\columnwidth]{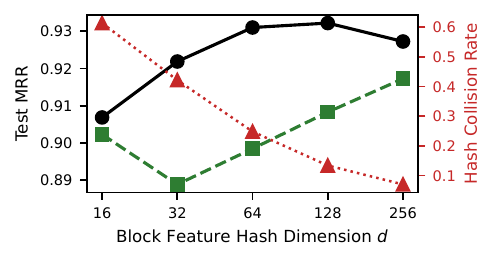}
\caption{Test MRR and empirical token hash collision rate across block feature hash dimension $d \in \{16, 32, 64, 128, 256\}$. $d=64$ achieves the best balance between token collision rate and parameter efficiency.}
\label{fig:hash_sens}
\end{figure}

We sweep the block feature hash dimension $d \in \{16,\allowbreak 32,\allowbreak 64,\allowbreak 128,\allowbreak 256\}$ and measure empirical token collision rates ($61.4\%$ for $d=16$, $42.2\%$ for $d=32$, $24.9\%$ for $d=64$, $13.4\%$ for $d=128$, and $7.0\%$ for $d=256$) alongside model accuracy (Fig.~\ref{fig:hash_sens}). GraphGPS MRR increases from $0.2342$ ($d=16$) and $0.2684$ ($d=32$) to $0.2894$ ($d=64$), with diminishing returns at $d=128$ ($0.2921$) and $d=256$ ($0.2935$) at $1.4\times$ memory cost. Sweeping GraphGPS block cap $c \in \{64,\allowbreak 128,\allowbreak 256,\allowbreak 512,\allowbreak \text{none}\}$ shows that $c=256$ ($2.48$~GB peak VRAM) retains the full accuracy of $c=512$ ($0.2901$ MRR, $4.39$~GB) and uncapped graphs ($0.2903$ MRR, $5.84$~GB). Table~\ref{tab:size_bins} breaks down performance across four function size bins ($1$--$16$, $17$--$64$, $65$--$256$, and $>256$ blocks), showing GraphGPS consistently outperforms GCN across all sizes.

\begin{table}[!t]
\caption{Retrieval Performance by Function Size Bin on the TREX Test Split}
\label{tab:size_bins}

\centering
\footnotesize
\setlength{\tabcolsep}{3.5pt}
\resizebox{\columnwidth}{!}{%
\begin{tabular}{@{}lccccc@{}}
\toprule
\textbf{Size Bin (Blocks)} & \textbf{Test Share} & \textbf{GCN MRR} & \textbf{GraphGPS MRR} & \textbf{GCN R@10} & \textbf{GraphGPS R@10} \\
\midrule
$1$--$16$ & $73.84\%$ & $0.2584 \pm 0.0014$ & $\mathbf{0.2942 \pm 0.0014}$ & $0.3912 \pm 0.0020$ & $\mathbf{0.4538 \pm 0.0020}$ \\

$17$--$64$ & $22.41\%$ & $0.2268 \pm 0.0024$ & $\mathbf{0.2815 \pm 0.0024}$ & $0.3495 \pm 0.0033$ & $\mathbf{0.4362 \pm 0.0033}$ \\

$65$--$256$ & $3.54\%$ & $0.1892 \pm 0.0062$ & $\mathbf{0.2468 \pm 0.0062}$ & $0.3042 \pm 0.0084$ & $\mathbf{0.3945 \pm 0.0084}$ \\

$>256$ & $0.21\%$ & $0.1425 \pm 0.0109$ & $\mathbf{0.1984 \pm 0.0109}$ & $0.2368 \pm 0.0149$ & $\mathbf{0.3270 \pm 0.0149}$ \\
\bottomrule
\end{tabular}%
}
\end{table}

\subsection{Reproducibility and Limitations}

Although Hex-Rays microcode provides strong normalization across compiler optimizations, its reliance on the commercial IDA Pro decompiler restricts headless extraction by users without a license. To ensure full reproducibility of our learning and evaluation pipeline, we release the complete source code, deterministic split definitions, pre-extracted microcode graphs, trained checkpoints, and table-generation scripts, allowing all experiments in Tables~\ref{tab:trex}--\ref{tab:size_bins} to be reproduced without IDA Pro. In addition, we provide an open-source Ghidra P-Code extraction front end (\texttt{extract/ghidra\_pcode.py}) that produces the exact same graph schema; on our \texttt{zlib} cross-optimization validation benchmark ($734$ functions across \texttt{-O0} through \texttt{-O3}), GCN trained on open front-end graphs reaches $0.9272 \pm 0.0133$ ROC-AUC and $0.7008 \pm 0.0825$ MRR. Remaining limitations include cross-architecture matching (e.g., x86-64 to ARM/MIPS) where architecture-specific calling conventions and intrinsic lowering still introduce variance, and heavy virtualization obfuscators that synthesize interpreter loops.

\section{Conclusion and Future Work}

This work presented MicroSim, a pipeline for binary code similarity detection based on decompiler microcode graphs and Siamese GNNs. On the TREX cross-optimization benchmark ($849{,}072$ function-level graphs spanning optimization levels \texttt{-O0} through \texttt{-O3}), evaluated under a strict class-disjoint split, MicroSim with a GraphGPS encoder reaches $0.9973$ ROC-AUC and $0.9756$ MCC in under ten minutes on a single laptop GPU, outperforming Gemini, SAFE, jTrans, and a structure-free BoW baseline. A multi-seed, compute-matched ablation demonstrates that the supervised-contrastive InfoNCE objective is preferable for ranked retrieval and thresholded classification (F1, MCC, MRR, and $\mathrm{R@}K$), while the pairwise cosine-embedding loss maintains a slightly higher raw ROC-AUC. Component and sensitivity ablations further confirm that CFG edges, DFG def--use edges, and 64-d hashed block features each contribute significantly to performance.

Several directions for future work stand out. Because single-GPU compute is not the primary bottleneck, further gains in large-pool retrieval will likely come from memory-queue hard-negative mining and cross-optimization view augmentations tailored to microcode. Extending the open Ghidra P-Code front end to full-scale multi-architecture corpora (x86-64, ARM64, and MIPS) will further broaden applicability in firmware and cross-platform vulnerability search.

\bibliographystyle{IEEEtran}
\bibliography{refs}

@article{haq2021survey,
  author    = {Haq, Irfan Ul and Caballero, Juan},
  title     = {A Survey of Binary Code Similarity},
  journal   = {ACM Comput. Surv.},
  volume    = {54},
  number    = {3},
  pages     = {51:1--51:38},
  month     = apr,
  year      = {2021},
  doi       = {10.1145/3446371}
}

@inproceedings{moussaoui2025bin2vec,
  author    = {Moussaoui, Moussa and Houichime, Tarik and Sadiq, Abdelalim},
  title     = {{Bin2Vec}: Interpretable and Auditable Multi-View Binary Analysis for Code Plagiarism Detection},
  booktitle = {Proc. Int. Conf. Adv. Commun. Technol. Netw. (CommNet)},
  address   = {Rabat, Morocco},
  month     = dec,
  year      = {2025},
  pages     = {1--8},
  doi       = {10.1109/CommNet68224.2025.11288883}
}

@inproceedings{xu2017gemini,
  author    = {Xu, Xiaojun and Liu, Chang and Feng, Qian and Yin, Heng and Song, Le and Song, Dawn},
  title     = {Neural Network-Based Graph Embedding for Cross-Platform Binary Code Similarity Detection},
  booktitle = {Proc. ACM SIGSAC Conf. Comput. Commun. Secur. (CCS)},
  address   = {Dallas, TX, USA},
  month     = oct,
  year      = {2017},
  pages     = {363--376},
  doi       = {10.1145/3133956.3134018}
}

@inproceedings{ding2019asm2vec,
  author    = {Ding, Steven H. H. and Fung, Benjamin C. M. and Charland, Philippe},
  title     = {{Asm2Vec}: Boosting Static Representation Robustness for Binary Clone Search Against Code Obfuscation and Compiler Optimization},
  booktitle = {Proc. IEEE Symp. Secur. Privacy (SP)},
  address   = {San Francisco, CA, USA},
  month     = may,
  year      = {2019},
  pages     = {472--489},
  doi       = {10.1109/SP.2019.00003}
}

@inproceedings{li2021palmtree,
  author    = {Li, Xuezixiang and Qu, Yu and Yin, Heng},
  title     = {{PalmTree}: Learning an Assembly Language Model for Instruction Embedding},
  booktitle = {Proc. ACM SIGSAC Conf. Comput. Commun. Secur. (CCS)},
  address   = {Virtual Event, Republic of Korea},
  month     = nov,
  year      = {2021},
  pages     = {3236--3251},
  doi       = {10.1145/3460120.3484587}
}

@inproceedings{massarelli2019safe,
  author    = {Massarelli, Luca and Di Luna, Giuseppe Antonio and Petroni, Fabio and Baldoni, Roberto and Querzoni, Leonardo},
  title     = {{SAFE}: Self-Attentive Function Embeddings for Binary Similarity},
  booktitle = {Proc. Int. Conf. Detection Intrusions Malware Vulnerability Assessment (DIMVA)},
  address   = {Gothenburg, Sweden},
  month     = jun,
  year      = {2019},
  pages     = {309--329},
  doi       = {10.1007/978-3-030-22038-9_15}
}

@inproceedings{pei2021trex,
  author    = {Pei, Kexin and Xuan, Zhou and Yang, Junfeng and Jana, Suman and Ray, Baishakhi},
  title     = {Trex: Learning Execution Semantics from Micro-Traces for Binary Similarity},
  booktitle = {Proc. IEEE Symp. Secur. Privacy (SP)},
  address   = {San Francisco, CA, USA},
  month     = may,
  year      = {2021},
  pages     = {1476--1492},
  doi       = {10.1109/SP40001.2021.00017}
}

@inproceedings{wang2022jtrans,
  author    = {Wang, Hao and Qu, Wenjie and Katz, Gilad and Zhu, Wenyu and Gao, Zeyu and Qiu, Han and Zhuge, Jianwei and Zhang, Chao},
  title     = {{jTrans}: Jump-Aware Transformer for Binary Code Similarity Detection},
  booktitle = {Proc. ACM SIGSOFT Int. Symp. Softw. Testing Anal. (ISSTA)},
  address   = {Virtual Event, Republic of Korea},
  month     = jul,
  year      = {2022},
  pages     = {1--13},
  doi       = {10.1145/3533767.3534367}
}

@inproceedings{feng2016genius,
  author    = {Feng, Qian and Zhou, Rundong and Xu, Chengcheng and Cheng, Yao and Testa, Brian and Yin, Heng},
  title     = {Scalable Graph-Based Bug Search for Firmware Images},
  booktitle = {Proc. ACM SIGSAC Conf. Comput. Commun. Secur. (CCS)},
  address   = {Vienna, Austria},
  month     = oct,
  year      = {2016},
  pages     = {480--491},
  doi       = {10.1145/2976749.2978370}
}

@inproceedings{li2019gmn,
  author    = {Li, Yujia and Gu, Chenjie and Dullien, Thomas and Vinyals, Oriol and Kohli, Pushmeet},
  title     = {Graph Matching Networks for Learning the Similarity of Graph Structured Objects},
  booktitle = {Proc. Int. Conf. Mach. Learn. (ICML)},
  address   = {Long Beach, CA, USA},
  month     = jun,
  year      = {2019},
  pages     = {3835--3845}
}

@inproceedings{duan2020deepbindiff,
  author    = {Duan, Yue and Li, Xuezixiang and Wang, Jinghan and Yin, Heng},
  title     = {{DeepBinDiff}: Learning Program-Wide Code Representations for Binary Diffing},
  booktitle = {Proc. Netw. Distrib. Syst. Secur. Symp. (NDSS)},
  address   = {San Diego, CA, USA},
  month     = feb,
  year      = {2020},
  pages     = {1--18},
  doi       = {10.14722/ndss.2020.24311}
}

@inproceedings{liu2018alphadiff,
  author    = {Liu, Bingchang and Huo, Wei and Zhang, Chao and Li, Wenchao and Li, Feng and Piao, Aihua and Zou, Wei},
  title     = {{$\alpha$Diff}: Cross-Version Binary Code Similarity Detection with {DNN}},
  booktitle = {Proc. IEEE/ACM Int. Conf. Automated Softw. Eng. (ASE)},
  address   = {Montpellier, France},
  month     = sep,
  year      = {2018},
  pages     = {667--678},
  doi       = {10.1145/3238147.3238199}
}

@inproceedings{yu2020ordermatters,
  author    = {Yu, Zeping and Cao, Rui and Tang, Qiyi and Nie, Sen and Huang, Junzhou and Wu, Shi},
  title     = {Order Matters: Semantic-Aware Neural Networks for Binary Code Similarity Detection},
  booktitle = {Proc. AAAI Conf. Artif. Intell. (AAAI)},
  address   = {New York, NY, USA},
  month     = feb,
  year      = {2020},
  pages     = {1145--1152},
  doi       = {10.1609/aaai.v34i01.5466}
}

@inproceedings{marcelli2022how,
  author    = {Marcelli, Andrea and Graziano, Mariano and Ugarte-Pedrero, Xabier and Fratantonio, Yanick and Mansouri, Mohamad and Balzarotti, Davide},
  title     = {How Machine Learning Is Solving the Binary Function Similarity Problem},
  booktitle = {Proc. USENIX Secur. Symp.},
  address   = {Boston, MA, USA},
  month     = aug,
  year      = {2022},
  pages     = {2099--2116}
}

@inproceedings{kipf2017gcn,
  author    = {Kipf, Thomas N. and Welling, Max},
  title     = {Semi-Supervised Classification with Graph Convolutional Networks},
  booktitle = {Proc. Int. Conf. Learn. Represent. (ICLR)},
  address   = {Toulon, France},
  month     = apr,
  year      = {2017},
  pages     = {1--14}
}

@inproceedings{hamilton2017graphsage,
  author    = {Hamilton, William L. and Ying, Rex and Leskovec, Jure},
  title     = {Inductive Representation Learning on Large Graphs},
  booktitle = {Proc. Adv. Neural Inf. Process. Syst. (NeurIPS)},
  address   = {Long Beach, CA, USA},
  month     = dec,
  year      = {2017},
  pages     = {1024--1034}
}

@inproceedings{velickovic2018gat,
  author    = {Veli{\v{c}}kovi{\'c}, Petar and Cucurull, Guillem and Casanova, Arantxa and Romero, Adriana and Li{\`o}, Pietro and Bengio, Yoshua},
  title     = {Graph Attention Networks},
  booktitle = {Proc. Int. Conf. Learn. Represent. (ICLR)},
  address   = {Vancouver, BC, Canada},
  month     = apr,
  year      = {2018},
  pages     = {1--12}
}

@inproceedings{xu2019gin,
  author    = {Xu, Keyulu and Hu, Weihua and Leskovec, Jure and Jegelka, Stefanie},
  title     = {How Powerful are Graph Neural Networks?},
  booktitle = {Proc. Int. Conf. Learn. Represent. (ICLR)},
  address   = {New Orleans, LA, USA},
  month     = may,
  year      = {2019},
  pages     = {1--17}
}

@inproceedings{gilmer2017mpnn,
  author    = {Gilmer, Justin and Schoenholz, Samuel S. and Riley, Patrick F. and Vinyals, Oriol and Dahl, George E.},
  title     = {Neural Message Passing for Quantum Chemistry},
  booktitle = {Proc. Int. Conf. Mach. Learn. (ICML)},
  address   = {Sydney, NSW, Australia},
  month     = aug,
  year      = {2017},
  pages     = {1263--1272}
}

@inproceedings{ying2021graphormer,
  author    = {Ying, Chengxuan and Cai, Tianle and Luo, Shengjie and Zheng, Shuxin and Ke, Guolin and He, Di and Shen, Yanming and Liu, Tie-Yan},
  title     = {Do Transformers Really Perform Bad for Graph Representation?},
  booktitle = {Proc. Adv. Neural Inf. Process. Syst. (NeurIPS)},
  address   = {Virtual Event},
  month     = dec,
  year      = {2021},
  pages     = {28877--28888}
}

@inproceedings{rampasek2022graphgps,
  author    = {Ramp{\'a}{\v{s}}ek, Ladislav and Galkin, Michael and Dwivedi, Vijay Prakash and Luu, Anh Tuan and Wolf, Guy and Beaini, Dominique},
  title     = {Recipe for a General, Powerful, Scalable Graph Transformer},
  booktitle = {Proc. Adv. Neural Inf. Process. Syst. (NeurIPS)},
  address   = {New Orleans, LA, USA},
  month     = nov,
  year      = {2022},
  pages     = {14501--14515}
}

@inproceedings{dwivedi2022graph,
  author    = {Dwivedi, Vijay Prakash and Luu, Anh Tuan and Laurent, Thomas and Bengio, Yoshua and Bresson, Xavier},
  title     = {Graph Neural Networks with Learnable Structural and Positional Representations},
  booktitle = {Proc. Int. Conf. Learn. Represent. (ICLR)},
  address   = {Virtual Event},
  month     = apr,
  year      = {2022},
  pages     = {1--25}
}

@misc{van2020cpc,
  author       = {van den Oord, Aaron and Li, Yazhe and Vinyals, Oriol},
  title        = {Representation Learning with Contrastive Predictive Coding},
  year         = {2018},
  howpublished = {arXiv:1807.03748},
  url          = {https://arxiv.org/abs/1807.03748}
}

@inproceedings{chen2020simclr,
  author    = {Chen, Ting and Kornblith, Simon and Norouzi, Mohammad and Hinton, Geoffrey},
  title     = {A Simple Framework for Contrastive Learning of Visual Representations},
  booktitle = {Proc. Int. Conf. Mach. Learn. (ICML)},
  address   = {Virtual Event},
  month     = jul,
  year      = {2020},
  pages     = {1597--1607}
}

@inproceedings{khosla2020supervised,
  author    = {Khosla, Prannay and Teterwak, Piotr and Wang, Chen and Sarna, Aaron and Tian, Yonglong and Isola, Phillip and Maschinot, Aaron and Liu, Ce and Krishnan, Dilip},
  title     = {Supervised Contrastive Learning},
  booktitle = {Proc. Adv. Neural Inf. Process. Syst. (NeurIPS)},
  address   = {Virtual Event},
  month     = dec,
  year      = {2020},
  pages     = {18661--18673}
}

@inproceedings{you2020graphcl,
  author    = {You, Yuning and Chen, Tianlong and Sui, Yongduo and Chen, Ting and Wang, Zhangyang and Shen, Yang},
  title     = {Graph Contrastive Learning with Augmentations},
  booktitle = {Proc. Adv. Neural Inf. Process. Syst. (NeurIPS)},
  address   = {Virtual Event},
  month     = dec,
  year      = {2020},
  pages     = {5812--5823}
}

@inproceedings{kingma2015adam,
  author    = {Kingma, Diederik P. and Ba, Jimmy},
  title     = {Adam: A Method for Stochastic Optimization},
  booktitle = {Proc. Int. Conf. Learn. Represent. (ICLR)},
  address   = {San Diego, CA, USA},
  month     = may,
  year      = {2015},
  pages     = {1--15}
}

@article{delong1988comparing,
  author    = {DeLong, Elizabeth R. and DeLong, David M. and Clarke-Pearson, Daniel L.},
  title     = {Comparing the Areas Under Two or More Correlated Receiver Operating Characteristic Curves: A Nonparametric Approach},
  journal   = {Biometrics},
  volume    = {44},
  number    = {3},
  pages     = {837--845},
  month     = sep,
  year      = {1988},
  doi       = {10.2307/2531595}
}

@article{holm1979simple,
  author    = {Holm, Sture},
  title     = {A Simple Sequentially Rejective Multiple Test Procedure},
  journal   = {Scand. J. Statist.},
  volume    = {6},
  number    = {2},
  pages     = {65--70},
  year      = {1979}
}

@inproceedings{junod2015ollvm,
  author    = {Junod, Pascal and Rinaldini, Julien and Wehrli, Johan and Michielin, Julie},
  title     = {Obfuscator-{LLVM} -- Software Protection for the Masses},
  booktitle = {Proc. IEEE/ACM Int. Workshop Softw. Protection (SPRO)},
  address   = {Florence, Italy},
  month     = may,
  year      = {2015},
  pages     = {3--9},
  doi       = {10.1109/SPRO.2015.10}
}

\end{document}